\documentclass{ws-ijmpa}

\begin{document}

\markboth{Joanna Przerwa}
{Search for bremsstrahlung radiation in quasi--free $np~\to~np \gamma$ reaction}
%{Instructions for Typing Manuscripts (Paper's Title)}

%%%%%%%%%%%%%%%%%%%%% Publisher's Area please ignore %%%%%%%%%%%%%%%
%
\catchline{}{}{}{}{}
%
%%%%%%%%%%%%%%%%%%%%%%%%%%%%%%%%%%%%%%%%%%%%%%%%%%%%%%%%%%%%%%%%%%%%

\title{SEARCH FOR BREMSSTRAHLUNG RADIATION IN QUASI-FREE $NP~\to~NP \gamma$ REACTIONS}
%BremINSTRUCTIONS FOR TYPESETTING MANUSCRIPTS\\
%USING \TeX\ OR \LaTeX\footnote{For the title, try not to use more than 
%3 lines. Typeset the title in 10 pt roman, uppercase and 
%boldface.}
%}

%\author{\footnotesize J.~PRZERWA, R.~CZY{\.Z}YKIEWICZ, M.~JANUSZ, L.~JARCZYK, B.~KAMYS, 
%P.~KLAJA, P.~MOSKAL, C.~PISKOR--IGNATOWICZ, J.~SMYRSKI }
%\address{Nuclear Physics Department, Jagellonian University,
%Cracow 30-059, Poland} 
%
%\author{\footnotesize D.~GRZONKA, K.~KILIAN, W.~OELERT, T.~SEFZICK, P.~WINTER, M.~WOLKE, P.~W{\"U}STNER}
%\address{IKP and ZEL Forschungszentrum J{\"u}lich, 
%D-52425 J{\"u}lich, Germany}
%
%\author{\footnotesize H.-H.~ADAM, A.~KHOUKAZ, N.~LANG, R.~SANTO, A.~T{\"A}SCHNER, J.P.~WESSELS}
%\address{Institut f{\"u}r Kernphysik, Universit{\"a}t M{\"u}nster, Wilhelm-Klemm-Str. 9
%M{\"u}nster, 48149, Germany}
%
%\author{\footnotesize A.~BUDZANOWSKI}
%\address{Institute of Nuclear Physics,
% Cracow 31-342, Poland}
%
%\author{\footnotesize T.~RO{\.Z}EK, M.~SIEMASZKO, W.~ZIPPER}
%\address{Institute of Physics, University of Silesia,
%Katowice 40-007, Poland}

\author{\footnotesize J.~PRZERWA$^\ast$, H.-H.~ADAM$^\ddagger$, A.~BUDZANOWSKI$^{\dagger \dagger}$, 
                      R.~CZY{\.Z}YKIEWICZ$^{\ast,\dagger}$, D.~GRZONKA$^\dagger$, M.~JANUSZ$^\ast$, 
                      L.~JARCZYK$^\ast$, B.~KAMYS$^\ast$, A.~KHOUKAZ$^\ddagger$,
                      K.~KILIAN$^\dagger$, P.~KLAJA$^\ast$, N.~LANG$^\ddagger$, P.~MOSKAL$^{\ast,\dagger}$, 
                      W.~OELERT$^\dagger$, C.~PISKOR--IGNATOWICZ$^\ast$,
                      T.~RO{\.Z}EK$^{\star,\dagger}$, R.~SANTO$^\ddagger$, T.~SEFZICK$^\dagger$,
                      M.~SIEMASZKO$^\star$, J.~SMYRSKI$^\ast$,
                      A.~T{\"A}SCHNER$^\ddagger$, P.~WINTER$^\dagger$, M.~WOLKE$^\dagger$, 
                      P.~W{\"U}STNER$^{\ddagger \ddagger}$,W.~ZIPPER$^\star$ }
\address{
$^\ast$ Nuclear Physics Department, Jagellonian University, Cracow, 30-059, Poland\\
$^\dagger$ Institut f{\"u}r Kernphysik, Forschungszentrum J{\"u}lich, J{\"u}lich, 52425, Germany\\
$^\ddagger$ Institut f{\"u}r Kernphysik, Universit{\"a}t M{\"u}nster,M{\"u}nster, 48149, Germany\\
$^{\dagger \dagger}$ Institute of Nuclear Physics, Cracow, 31-342, Poland\\
$^\star$ Institute of Physics, University of Silesia, Katowice, 40-007, Poland\\
$^{\ddagger \ddagger}$ ZEL Forschungszentrum J{\"u}lich, J{\"u}lich, 52425, Germany\\
}

\maketitle

\pub{Received (Day Month Year)}{Revised (Day Month Year)}

\begin{abstract}
   Due to the high sensitivity of the $NN~\to~NN\gamma$
   reaction to the nucleon~--~nucleon potential, Bremsstrahlung radiation is used
   as a tool to investigate details of the nucleon~--~nucleon interaction.
   Such investigations
   can be performed at the cooler synchrotron COSY in the Research Centre
   J{\"u}lich, by dint of the COSY--11 detection system.\\
   The results of the identification of Bremsstrahlung radiation
   emitted via the $dp~\to~dp\gamma$ reaction
   in data taken with a proton target and a deuteron beam
   are presented and discussed.\\

\vspace{-0.4cm}


\end{abstract}

\vspace{0.3cm}

The installation of a neutron detector at the COSY-11 facility~[1,2]
enables to study a plethora of new reaction channels.
It opens wide possibilities not only to investigate the isospin dependence of the
meson production~[3],
but also to measure the Bremsstrahlung radiation
created in the collisions of nucleons.
The study of the latter process is interested since it is
highly sensitive to the kind of the nucleon-nucleon potential,
and hence may
serve as a tool  to discriminate between various existing potential
models~[4,5]. Although Bremsstrahlung radiation has been studied since many years,
it is still the subject of interest of many theoretical and experimental groups~[4,5,6,7]. \\
At the COSY--11 experiment a signal from  $\gamma$--quanta was observed
in the time--of--flight distribution for the neutral particles measured between the target
and the neutral particle detector~[2]. This encouraged us to analyse the data in view of the 
Bremsstrahlung radiation in a free $dp~\to~dp \gamma$ and a quasi--free $np~\to~np \gamma$ reactions.
\begin{figure}[t]
\parbox{0.45\textwidth}{\psfig{file=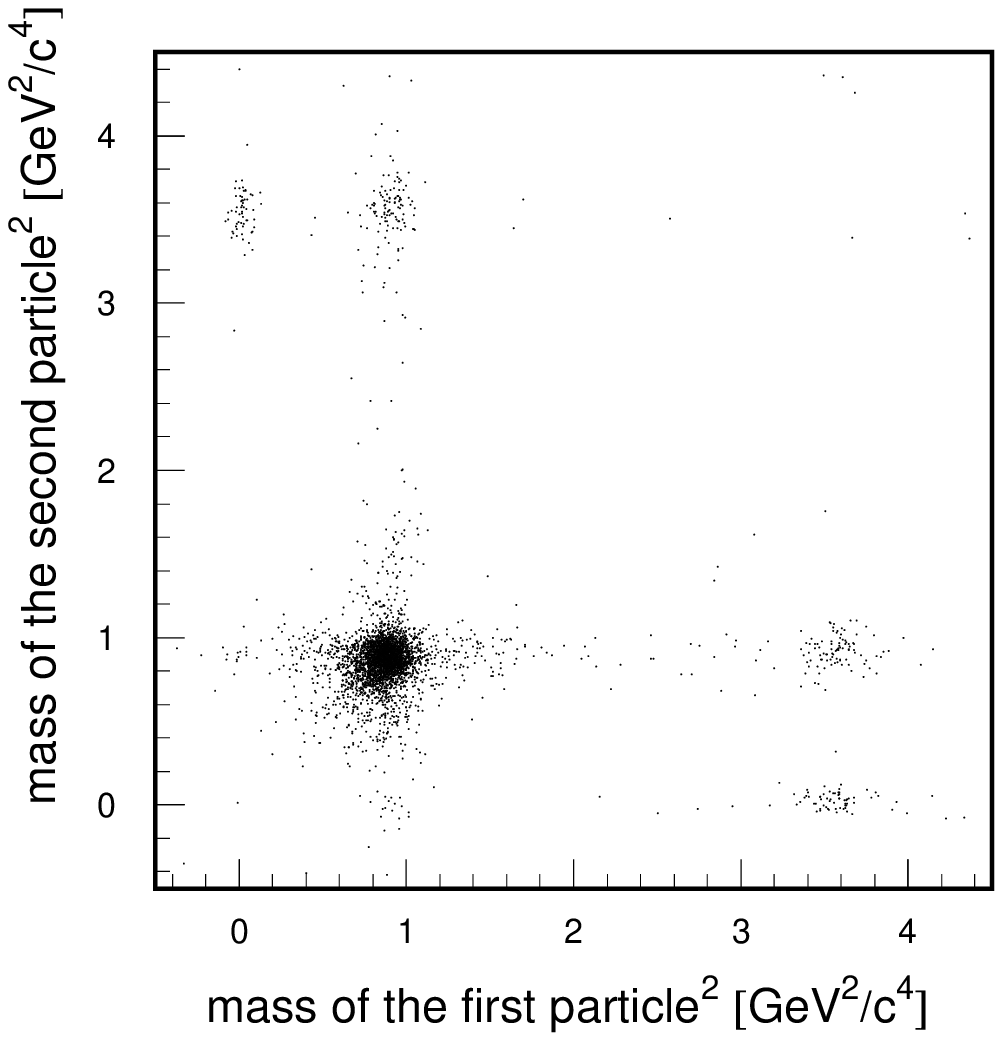,width=6cm}}
\parbox{0.45\textwidth}{\psfig{file=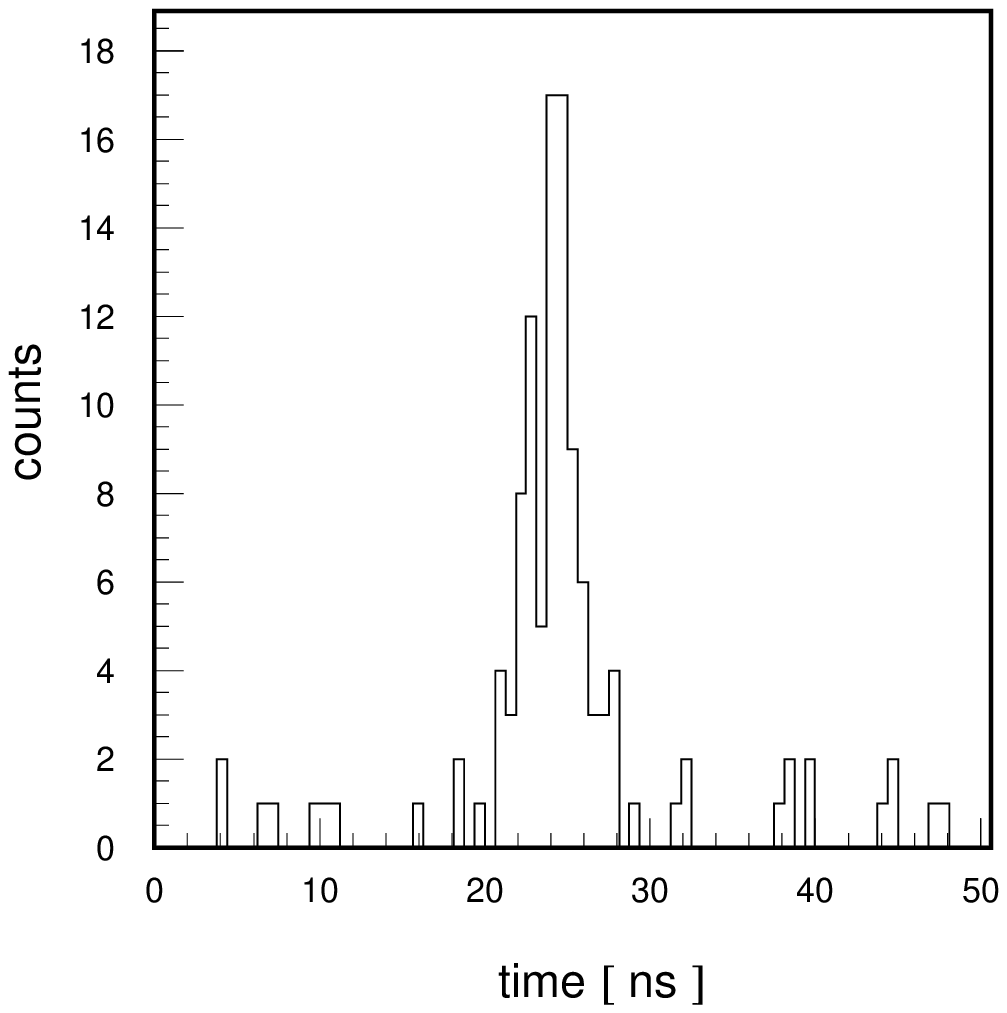,width=6cm}}
\vspace{-1.2cm}
\vspace*{8pt}
\caption{{\bf (left)} Scatter plot of invariant masses determined for events with two charged 
                      particles measured in coincidence. 
        {\bf (right)} Time--of--flight distribution obtained under assumption that
                      additionally to a signal an in the neutron detector,
                      one proton and one deuteron were identified.}
\end{figure}
Data have been taken using a proton target and a deuteron beam with a momentum close to the threshold
of the $dp~\to~dp \eta$ process. Events corresponding to the $dp~\to~dp \gamma$ and
$dp~\to~ppn \gamma$ reaction have been
identified by measuring the outgoing charged as well as neutral ejectiles.
The protons and deuterons are
detected by means of drift
chambers and scintillator hodoscopes while
neutrons and photons are
registered in a scintillator--lead sandwich type detector.
The momentum vectors of the charged ejectiles are  reconstructed
by tracking back the  trajectories to the target point~[1].
In case of a neutron the time--of--flight between the target and the neutral--particle--detector
together with the known position of the  hitted detection unit enables to determine its four--momentum vector~[8].\\
In order to identify the $dp~\to~dp \gamma$ reaction events with two tracks
in the drift chambers and a simultaneous signal in the neutron detector have been
selected. In fig.1 (left) the squared mass of one particle is plotted versus 
the squared mass of the other registered particle. Based on this figure the measured reactions 
can be grouped according to the type of ejectiles. Thus reactions with two protons, 
proton and pion, proton and deuteron, and pion and deuteron can be clearly selected.
Next the distribution of the time--of--flight between the target 
neutron detector was determined with requirement that one of the charged particles was
identified as a proton and the other as a deuteron. Due to the baryon number 
conservation,  gamma quanta are the only one possible source of a signal in a neutron
detector. Indeed a clear peak around the time corresponding to the time--of--light
of the light is visible (see fig.1 right). The gamma quanta may originate from Bremsstrahlung reaction
or from the decay of produced mesons eg. via the $dp~\to~dp \pi^{0}~\to~dp \gamma \gamma$.
It is possible to distinguish between these hypothesis calculating the missing mass produced
in the $dp~\to~dpX$ reaction. Fig.2 left shows the distribution of the squared missing mass as obtained
for the $dp~\to~dpX$ reaction. A significant peak around 0~MeV$^2$/c$^4$ --- the squared
mass of a gamma quanta --- constitute an evidence for events associated to the deuteron--proton
bremsstrahlung. In addition a broad structure at higher masses originating from two pions emmited
from reaction $dp~\to~dp \pi^{0} \pi^{0}$ or two gamma quanta from $dp~\to~dp \gamma \gamma$ reaction is visible.
The extraction of the total cross section 
of the $dp~\to~dp \gamma$ reaction requires the luminosity and acceptance determination,
which will be performed in the near future.
The analysis of the $dp~\to~np \gamma p_{sp}$ reaction is more 
complicated, however in this case all three baryons, namely two protons
and neutron can be measured in the drift chambers and in the neutron detector, respectively.
Fig.2 (right) presents the time--of--flight with the condition that in coincidence with 
a neutral particle also two protons were identified. A clear signal originating from
gamma quanta is seen at a time value of 24~ns which coresponds to the velocity of light. 
Using the missing mass technique, it might be possible to verify if gamma quanta originate from the direct
$dp~\to~np \gamma p_{sp}$ reaction. At present the data analysis is  in progress.

%\begin{figure}[H]
%\parbox{0.45\textwidth}{\psfig{file=mass_vs_mass.eps,width=5cm}}
%\parbox{0.45\textwidth}{\psfig{file=tof_dp_dpgamma.eps,width=5cm}}
%\vspace{-1.2cm}
%\vspace*{8pt}
%\caption{{\bf (left)} Scatter plot of invariant masses determined for events with two charged 
%                      particles measured in coincidence. 
%        {\bf (right)} Time--of--flight distribution obtained under assumption that
%                      additionally to a signal an in the neutron detector,
%                      one proton and one deuteron were identified.}
%\end{figure}

\begin{figure}
\parbox{0.45\textwidth}{\psfig{file=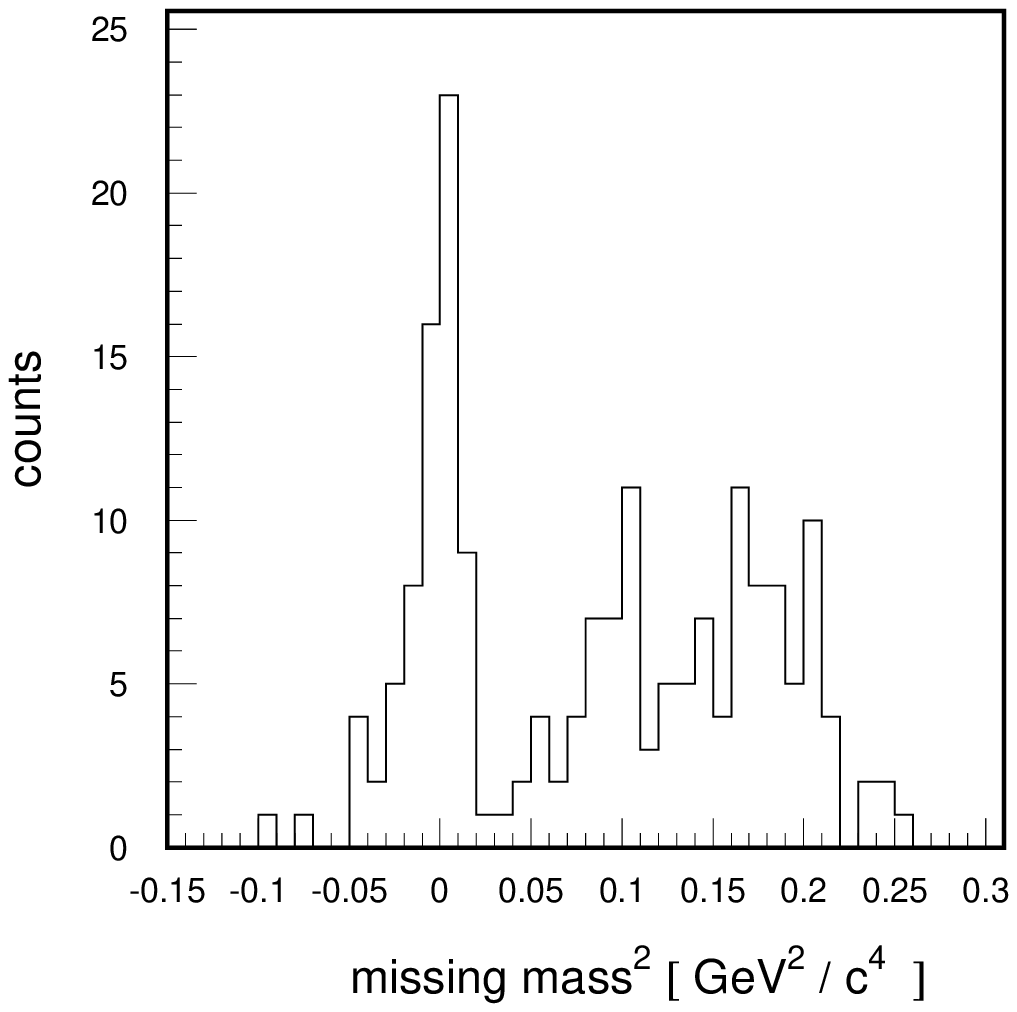,width=6cm}}
\parbox{0.45\textwidth}{\psfig{file=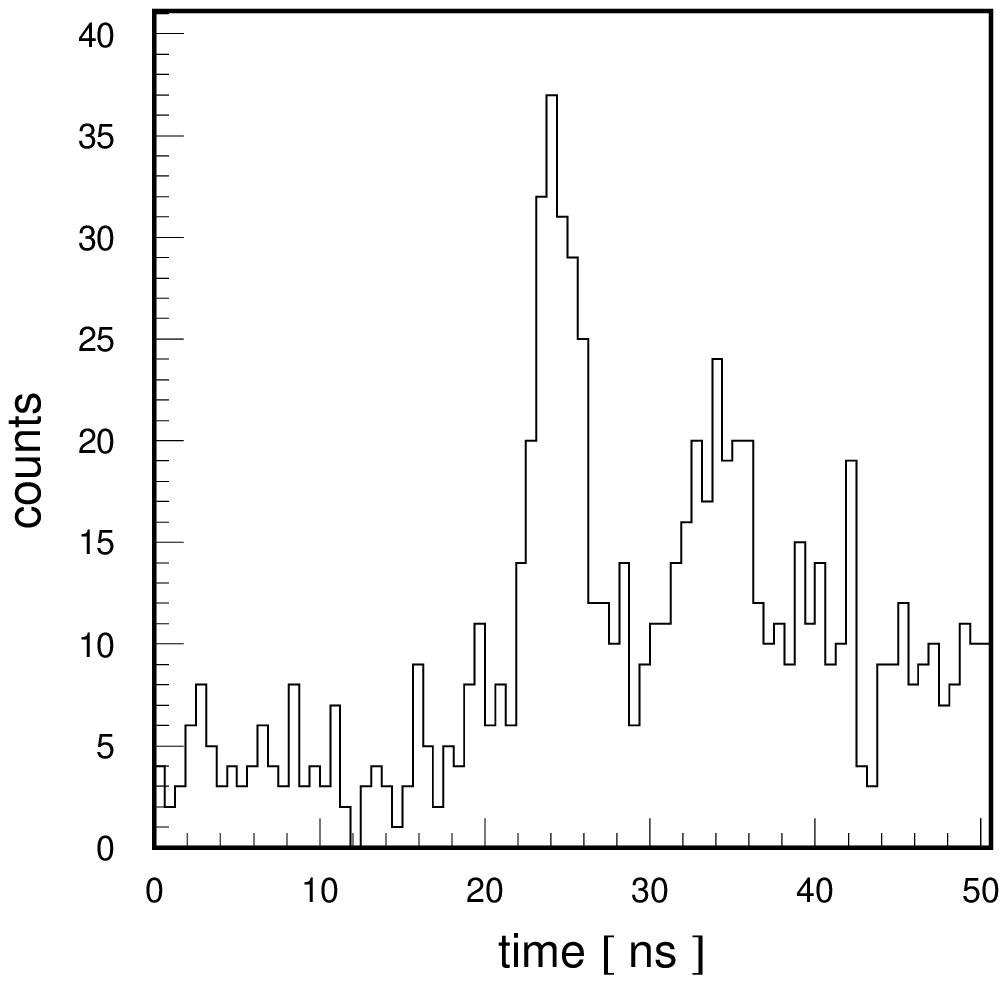,width=6cm}}
\vspace{-1.2cm}
\vspace*{8pt}
\caption{{\bf (left)} Distribution of the squared missing
                      mass for the $dp~\to~dpX$ reaction.
         {\bf (right)}Time--of--flight distribution obtained under assumption that 
                      additionally to a signal an in the neutron detector, 
                      two protons were identified.}
\end{figure}

\vspace{-0.4cm}

\section*{Acknowledgments}

 The work has been supported by the European Community - Access to
 Research Infrastructure action of the
 Improving Human Potential Programme,
 by the DAAD Exchange Programme (PPP-Polen),
 by the Polish State Committe for Scientific Research
 (grants No. 2P03B07123 and PB1060/P03/2004/26)
 and by the Research Centre J{\"u}lich.
%\appendix

%\section{Appendices}

%\section*{References}

\end{document}